\documentclass[10pt, prb, aps, twocolumn, showpacs, citeautoscript, floatfix, reprint, amsmath, amssymb, notitlepage, superscriptaddress]{revtex4-1}

\usepackage{booktabs}
\usepackage[utf8]{inputenc}
\usepackage[T1]{fontenc}
\usepackage[english]{babel}
\usepackage{graphicx}
\usepackage{rotating}
\usepackage{mathtools}
\usepackage{amsfonts}
\usepackage{dcolumn} 
\usepackage{bm} 
\usepackage{xcolor}
\usepackage{latexsym}
\usepackage{wasysym}
\usepackage[colorlinks, citecolor=blue, linkcolor=blue, urlcolor=blue]{hyperref}
\usepackage[nomain, acronym, shortcuts]{glossaries}
\usepackage{tikz}
\usepackage{environ}
\usepackage{csquotes}
\usepackage[normalem]{ulem}
\usepackage{amsmath}
\DeclareMathOperator{\Tr}{Tr}

\glsdisablehyper
\usetikzlibrary{external}
\bibpunct{[}{]}{,}{n}{}{}

\begin{document}

\title{Corroborating the bulk-edge correspondence in weakly interacting 1D
topological insulators}



\author{Antonio Zegarra}

\affiliation{Department of Physics, PUC-Rio, 22451-900 Rio de Janeiro, Brazil}

\author{Denis R. Candido}
\affiliation{Instituto de F\'{i}sica de S\~ao Carlos, Universidade de S\~ao Paulo, 13560-970 S\~ao Carlos, S\~ao Paulo, Brazil}
\affiliation{Institute for Molecular Engineering, University of Chicago, Chicago, Illinois 60637, USA}
\affiliation{Department of Physics and Astronomy and Optical Science and Technology Center, University of Iowa, Iowa City, Iowa 52242, USA}

\author{J. Carlos Egues}
\affiliation{Instituto de F\'{i}sica de S\~ao Carlos, Universidade de S\~ao Paulo, 13560-970 S\~ao Carlos, S\~ao Paulo, Brazil}

\author{Wei Chen}

\affiliation{Department of Physics, PUC-Rio, 22451-900 Rio de Janeiro, Brazil}

\date{\today}

\begin{abstract}

We present a Green's function formalism to investigate the topological properties of weakly interacting one-dimensional topological insulators, including the bulk-edge correspondence and the quantum criticality near topological phase transitions, and using interacting Su-Schrieffer-Heeger model as an example. From the many-body spectral function, we find that closing of the bulk gap remains a defining feature even if the topological phase transition is driven by interactions. The existence of edge state in the presence of interactions can be captured by means of a $T$-matrix formalism combined with Dyson's equation, and the bulk-edge correspondence is shown to be satisfied even in the presence of interactions. The critical exponent of the edge state decay length is shown to be affiliated with the same universality class as the noninteracting limit.

\end{abstract}

\maketitle

\section{Introduction}

Topological order in noninteracting topological insulators (TIs) and topological superconductors (TSCs) is often calculated from the single-particle Bloch wave function. Depending on the dimension and symmetry class of the system\cite{Schnyder08,Kitaev09,Ryu10}, the index that characterizes the topology, namely the topological invariant ${\cal C}$, is calculated by different means (see, for instance, Ref.~\onlinecite{Chen19} for a summary). As the bulk gap $M$ of the system is varied, the topological invariant ${\cal C}$ remains the same integer, until the system reaches the critical point $M_{c}=0$ where ${\cal C}$ jumps abruptly, signaling a topological phase transition. As a consequence, the critical point $M_{c}=0$ is accompanied by a gap-closing in the bulk bands. The phases with zero and nonzero ${\cal C}$ are referred to as the topologically trivial and nontrivial phases, respectively, with the topologically nontrivial phase being of great interest in the literature. This is due to the existence of the edge state in the topologically nontrivial phase, known as the bulk-edge correspondence, which may have practical applications such as in non-Abelian quantum computation\cite{Nayak08} or spintronic devices\cite{Pesin12_2}. Interestingly, edge states have also been reported in topologically trivial quantum dots and strips ~\cite{candido2018prl,candido2018}. In all of those noninteracting systems, the edge state can be easily identified by solving the low energy Dirac Hamiltonian projected to real space\cite{Konig08,Zhou08,Linder09,Lu10,Shen12,Qi11}. In addition, if the TI or TSC exhibits linear band-crossing at the topological phase transition, the decay length of the edge state is inversely proportional to the bulk gap $\xi\propto|M|^{-1}$, yielding a critical exponent $\nu=1$.

Although the above unified picture nicely summarizes practically all noninteracting TIs and TSCs with linear band-crossing, the situation is very different when many-body interactions are included. In the presence of interactions, the single-particle Bloch state is no longer a valid quantity to start with, and hence the topology must be defined by other means. Limiting our discussions to weakly interacting TIs and TSCs, the many-body Green's function serves as a viable quantity to identify topology\cite{So85,Niu85,Ishikawa86,Volovik88,Volovik03,Qi08,Wang10,Gurarie11,
Essin11,Wang12,Wang12_2,Wang13,Grandi15}. Depending on the dimension and symmetry of the many-body Hamiltonian, the topological invariant ${\cal C}$ takes the form of the momentum space integration of a certain combination of Green's function and its derivative, and the effect of interactions is included in the self-energy that enters the Green's function\cite{Chen18}. Moreover, a curvature renormalization group (CRG) approach has been proposed to circumvent the tedious integration of the topological invariant, which has been proved as an efficient tool to identify topological phase transitions in a large parameter space\cite{Chen16,Chen16_2,Chen17,Chen18,Chen19}. 


Despite the success of the Green's function formalism, many issues concerning the weakly interacting TIs and TSCs await clarification. The first is whether the feature of gap-closing still manifests in interaction-driven topological phase transitions, since a reliable way to identify the bulk gap in the presence of interactions is required. The second is whether the bulk-edge correspondence remains valid. This is of significant importance, since it is crucial to clarify whether the edge state survives interactions, which is a realistic issue in practical applications. Finally, concerning the quantum criticality of the edge state, it is of fundamental interest to clarify whether the edge state decay length exhibits the same critical exponent as the noninteracting counterpart, because it answers whether the system remains in the same universality class as the noninteracting system when the interactions are included\cite{Chen17,Chen19}.

In this article, we present a perturbative formalism to clarify these issues. Limiting our discussions to one-dimensional (1D) TIs with electron-electron interaction, we first use the spectral function calculated out of the many-body Green's function to identify the feature of gap-closing near the topological phase transitions. Secondly, we demonstrate the validity of the bulk-edge correspondence in the presence of interactions using the many-body Green's function combined with the $T$-matrix formalism. The method essentially treats the edge as an impurity, thus breaking the periodicity and simulating the open boundary case. The edge state in this case is identified from the local density of states (LDOS) near the edge, whose decay length helps to extract the critical exponent.



\section{1D topological insulators with electron-electron interaction}

\subsection{Homogeneous Dirac models with interactions\label{sec:Green_fn_Dirac_interaction}}



Our goal here is to use the Green's function formalism to study 2$\times 2$ Dirac models in the presence of many-body interactions for the open boundary case. We start from the most general noninteracting 1D spinless 2$\times 2$ Dirac Hamiltonian
\begin{eqnarray}
&&{\cal H}_{0}=\sum_{k}
\left(
\begin{array}{cc}
c_{Ak}^{\dag} & c_{Bk}^{\dag} \\
\end{array}
\right)H_{0}(k)
\left(
\begin{array}{c}
c_{Ak} \\
c_{Bk}
\end{array}
\right)\;,
\nonumber \\
&&H_{0}(k)=-d_{0}(k)\sigma_{0}+d_{1}(k)\sigma_{1}+d_{2}(k)\sigma_{2}+d_{3}(k)\sigma_{3}\;,
\label{general_2by2_Dirac_Hamiltonian}
\end{eqnarray}
where $c_{Ik}^{\dagger}$ $(c_{Ik})$ with $\left\{c_{Ik},c_{I^{\prime}k^{\prime}}^{\dag}\right\}=\delta_{II^{\prime}}\delta_{kk^{\prime}}$ denotes the creation (annihilation) operator of spinless fermion with momentum $k$ and $I=A,B$ the sublattice degree of freedom. The presence of each $d_i(k)$ and their evenness and oddness in $k$ are determined by the Hamiltonian symmetry class. In this work we will focus on the BDI class, which in general can host topological trivial and non-trivial edge sates, mainly due the presence of the chiral symmetry~\cite{kharitonov2017,candido2018, candido2018prl, yao2009}. This symmetry class requires time reversal $({\mathcal T}^2=+1)$, particle-hole $({\mathcal C}=+1)$ and chiral $({\mathcal S})$ symmetries~\cite{Schnyder08,Ryu10,Chen19}, which constrains $d_i(k)$ such that 
\begin{align}
&d_{0}\left(k\right)=d_{3}\left(k\right)=0,\;\;\;
d_{1}\left(k\right)=d_{1}\left(-k\right),
\nonumber \\
&d_{2}\left(k\right)=-d_{2}\left(-k\right),
\label{d0}
\end{align}
To address the effect of weak electron-electron interaction, and the open boundary simulated by an impurity, we first solve the interaction-dressed Green's function $G_{int}$ from Dyson's equation. Subsequently, we use $G_{int}$ to solve for the full Green's function $G$ by means of the T-matrix approximation that incorporates the effect of the single impurity (the edge of the system).

Following the recipe above, we first calculate perturbatively the effect of the weak interaction Hamiltonian ${\cal H}_{int}$ using the Matsubara Green's function
\begin{eqnarray}
G_{int}(k,\tau)=\left(
\begin{array}{ll}
G_{intAA}(k,\tau) & G_{intAB}(k,\tau) \\
G_{intBA}(k,\tau) & G_{intBB}(k,\tau)
\end{array}
\right)\;,
\end{eqnarray}
with the matrix elements defined by 
\begin{eqnarray}
G_{intIJ}(k,\tau)=-\langle T_{\tau}c_{Ik}(\tau)c_{Jk}^{\dag}(0)\rangle\;,
\end{eqnarray}
where $T_{\tau}$ is the time ordering operator. The interaction-dressed Green's function with discrete frequency $i\omega_{n}$ can be obtained from Dyson's equation
\begin{eqnarray}
G_{int}&=&G_{0}+G_{0}\Sigma G_{int}=G_{0}+G_{0}\Sigma G_{0}+G_{0}\Sigma G_{0}\Sigma G_{0}+...
\nonumber \\
&=&\left(G_{0}^{-1}-\Sigma\right)^{-1}=\left(i\omega_{n}-H_{0}-\Sigma\right)^{-1},
\label{Dyson_equation}
\end{eqnarray}
where $\Sigma$ is the self-energy. The interacting part of Dyson's equation in Eq.~(\ref{Dyson_equation}) is
\begin{eqnarray}
&&\left(G_{0}\Sigma G_{int}\right)_{IJ}=-\sum_{n=1}^{\infty}(-1)^{n}\int_{0}^{\beta}d\tau_{1}\int_{0}^{\beta}d\tau_{2}...
\int_{0}^{\beta}d\tau_{n}
\nonumber \\
&&\times\langle T_{\tau} c_{Ik}(\tau){\cal H}_{int}(\tau_{1}){\cal H}_{int}(\tau_{2})...{\cal H}_{int}(\tau_{n})c_{Jk}^{\dag}(0)\rangle \label{eq9},
\end{eqnarray}
which corresponds to different connected diagrams. In this article, we restrict our calculation to one-loop level. In addition, we examine the short range electron-electron interaction between spinless fermions\cite{Wen10,Grushin13,Jia13,Daghofer14,Guo14,Luo15}. For TI, we consider the density-density interaction between two sublattices that takes the form
\begin{eqnarray}
{\cal H}_{e-e}=\sum_{kk^{\prime}q}V_{q}c_{A{k+q}}^{\dag}c_{B k^{\prime}-q}^{\dag}c_{Bk^{\prime}}c_{Ak}\;, 
\label{general_density_density_interaction}
\end{eqnarray}
where the precise form of $V_{\bf q}$ depends on the interactions considered in real space. 
\begin{figure}[ht]
\begin{center}
\includegraphics[clip=true,width=0.99\columnwidth]{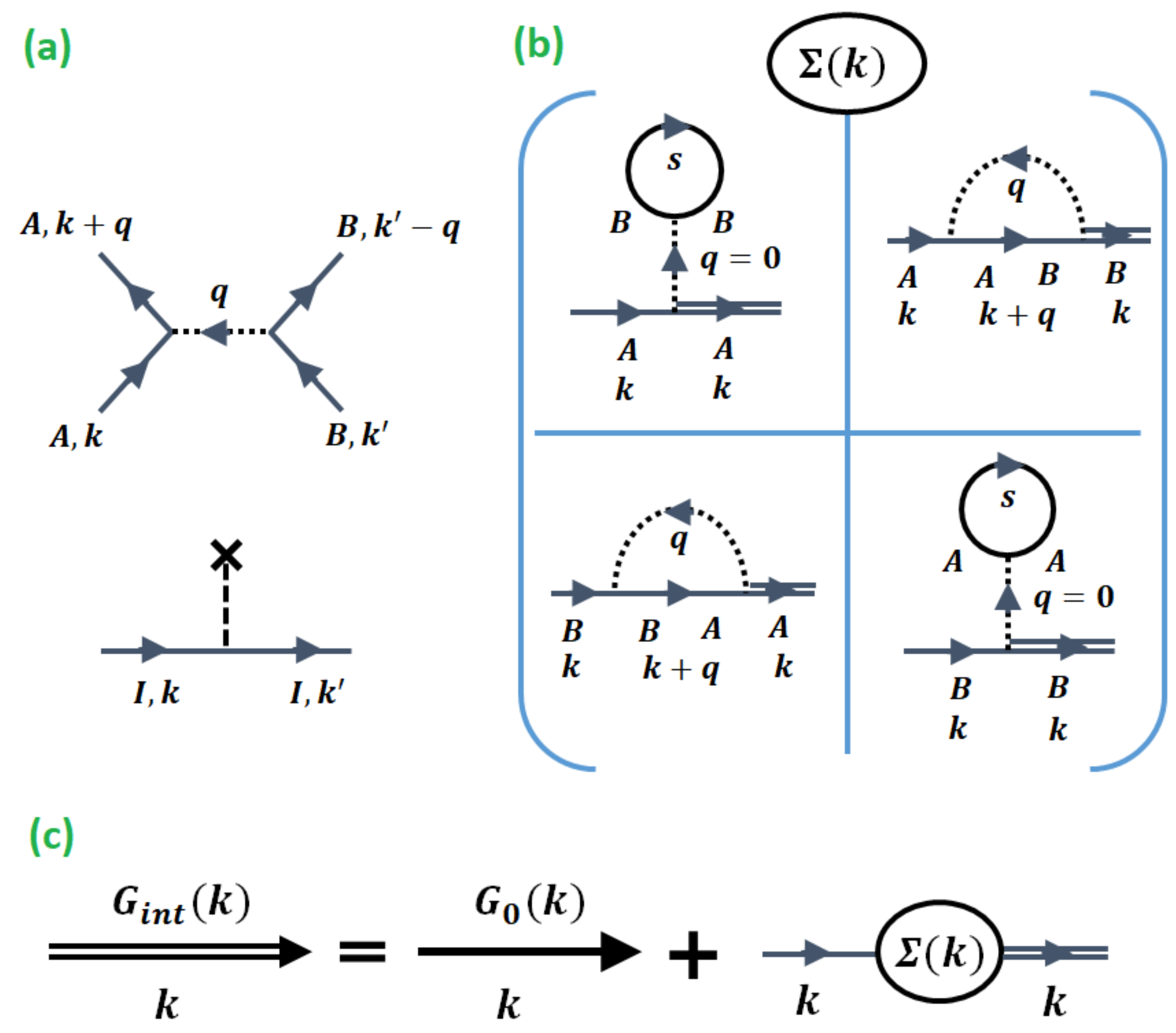}
\caption{ (a) The vertex for density-density interaction between sublattice $A$ and $B$ considered in this article, and the vertex of impurity scattering between the same sublattice. (b) The self-energy matrix for the density-density interaction case calculated up to one-loop Hartree-Fock level. (c) The Dyson's equation for the homogeneous interacting Green's function $G_{int}$. } 
\label{fig:self_energy_Feymann_diagram}
\end{center}
\end{figure}
Substituting Eq.~(\ref{general_density_density_interaction}) into (\ref{eq9}), the interacting part of the Dyson's equation up to one-loop level becomes
\begin{eqnarray}
&&\left(G_{0}\Sigma G_{int}\right)_{IJ}=\int_{0}^{\beta}d\tau_{1}\sum_{pp^{\prime}q}V_{q}
\nonumber \\
&&\times\langle T_{\tau}c_{Ik}(\tau)c_{Ap+q}^{\dag}(\tau_{1})c_{Bp^{\prime}-q}^{\dag}(\tau_{1})
c_{Bp^{\prime}}(\tau_{1})c_{Ap}(\tau_{1})c_{Jk}^{\dag}(0)\rangle.
\nonumber \\
\label{GSG_one_loop}
\end{eqnarray}
After Fourier transforming Eq.~(\ref{GSG_one_loop}), we obtain the self-energies (depicted in Fig.~\ref{fig:self_energy_Feymann_diagram} (b))
\begin{eqnarray}
\Sigma_{AA}(k)&=&\sum_{p}V_{q=0}G_{0BB}(p,\tau=0)\;,
\nonumber \\
\Sigma_{AB}(k)&=&-\sum_{q}V_{q}G_{0AB}(k+q,\tau=0)\;,
\nonumber \\
\Sigma_{BA}(k)&=&-\sum_{q}V_{q}G_{0BA}(k+q,\tau=0)\;,
\nonumber \\
\Sigma_{BB}(k)&=&\sum_{p}V_{q=0}G_{0AA}(p,\tau=0)\;.
\label{Hartree_Fock_self_energy}
\end{eqnarray}
Notice that the self-energies are frequency-independent $\Sigma_{IJ}(k,i\omega_{n})=\Sigma_{IJ}(k)$. 
After solving Dyson's equation we analytically continue its solution $i\omega_{n}\rightarrow i\omega$ to obtain the Green's function in terms of a continuous frequency $\omega$ from which we can determine the LDOS and the topological invariant as we explain in the next sections. The full Green's function in momentum-frequency space takes the form
\begin{eqnarray}
G_{int}(k,i\omega)
&=&\frac{1}{(i\omega+d_{0}^{\prime})^{2}-d^{\prime 2}}
\nonumber \\
&&\times\left(
\begin{array}{cc}
i\omega+d_{0}^{\prime}+d_{3}^{\prime} & d_{1}^{\prime}-id_{2}^{\prime} \\
d_{1}^{\prime}+id_{2}^{\prime} & i\omega+d_{0}^{\prime}-d_{3}^{\prime}
\end{array}
\right)\;,
\nonumber \\
G^{-1}_{int}(k,i\omega)&=&\left(
\begin{array}{cc}
i\omega+d_{0}^{\prime}-d_{3}^{\prime} & -d_{1}^{\prime}+id_{2}^{\prime} \\
-d_{1}^{\prime}-id_{2}^{\prime} & i\omega+d_{0}^{\prime}+d_{3}^{\prime}
\end{array}
\right)\;,
\label{2D_G_dprime}
\end{eqnarray}
where $d^{\prime}=\sqrt{d_{1}^{\prime 2}+d_{2}^{\prime 2}+d_{3}^{\prime 2}}$, and the self-energy-renormalized ${\bf d}^{\prime}$-vector ${\bf d}^{\prime}=\left(d_{1}^{\prime},d_{2}^{\prime},d_{3}^{\prime} \right)$ is
\begin{eqnarray}
&&d_{1}^{\prime}=d_{1}+{\rm Re}\Sigma_{AB}\;,\;\;\;d_{2}^{\prime}=d_{2}-{\rm Im}\Sigma_{AB}\;,
\nonumber \\
&&d_{3}^{\prime}=d_{3}+\frac{\Sigma_{AA}-\Sigma_{BB}}{2}\;,
\;\;\;d_{0}^{\prime}=\frac{-\Sigma_{AA}-\Sigma_{BB}}{2}\;.
\label{renormalized_d_vector_mu}
\end{eqnarray}
It is important to notice that because the self-energy $\Sigma_{IJ}=\Sigma_{IJ}(k)$ is a function of only $k$, so is each $d_{i}^{\prime}=d_{i}^{\prime}(k)$.


We also calculate the spectral function through the retarded version of the Green's function Eq.~(\ref{2D_G_dprime}) with real frequency $i\omega\rightarrow\omega+i\eta$ (analytical continuation), defined by 
\begin{eqnarray}
A(k,\omega)=-\frac{1}{\pi}{\rm Im}\left[{\rm Tr}G_{int}^{ret}(k,\omega)\right]\;.
\label{spectral_fn_Gint}
\end{eqnarray}
As we discuss in the next sections, the spectral function is the key to identify the bulk gap and the gap-closing (topological phase transitions) in the presence of interaction.



It is important to emphasize that the interaction term ${\cal H}_{e-e}$ in Eq.~(\ref{general_density_density_interaction}) breaks both chiral and particle-hole symmetries, thus changing the symmetry of the effective Hamiltonian ${H}(k)=-\sigma_0{d_0}^{\prime}+\boldsymbol{\sigma} \cdot \bf{d}^{\prime}$. 
More specifically, in the case of density-density interaction in which the self-energy is frequency-independent, we find $d_{3}^{\prime}=0$ and a finite $d_0^{\prime}$ (See Sec.~III for details). 
Although the standard definition of $\mathcal{S}$ and $\mathcal{C}$\cite{Ryu10} requires Eq.~(\ref{d0}) -- incompatible with having a finite $d_0^{\prime}$ -- Ref.~\onlinecite{Schnyder08} noticed the possibility of generalizing those symmetries with respect to a shift. With this generalization a system that has both particle-hole and chiral symmetries satisfies 
\begin{eqnarray}
&&{\mathcal S}\left[H\left(k\right)-\frac{1}{2}\Tr H\left(k\right)\right]{\mathcal S}^{-1}=-\left[H\left(k\right)-\frac{1}{2}\Tr H\left(k\right)\right]
\nonumber \\
&&={\mathcal C}\left[H\left(-k\right)-\frac{1}{2}\Tr H\left(-k\right)\right]{\mathcal C}^{-1}.
\label{newC}
\end{eqnarray}
Accordingly, the entire spectrum is shifted by $\Tr H(k)/2=-d_{0}^{\prime}$ and hence the spectrum can be thought to be particle-hole and chiral symmetric with respect to $-d_{0}^{\prime}$. 
The noninteracting parametrization in Eq.~(\ref{d0}) is the special case when $\Tr H(k)=0$.

\subsection{Edge density of states of interacting 1D TIs \label{sec:edge_state_Dirac}}

We now address the edge state in the presence of interactions. Because of the interactions, identifying edge state from the single particle wave function is no longer feasible. Instead, the valid quantity to be investigated is the impurity Green's function, as has been demonstrated for noninteracting systems\cite{Slager15}. Our purpose is to further demonstrate that the LDOS calculated out of the real space impurity Green's function, together with the topological invariant calculation, can be used to identify the bulk-edge correspondence and quantum criticality for interacting 1D TIs. 

\begin{figure}[ht]
\begin{center}
\includegraphics[clip=true,width=0.99\columnwidth]{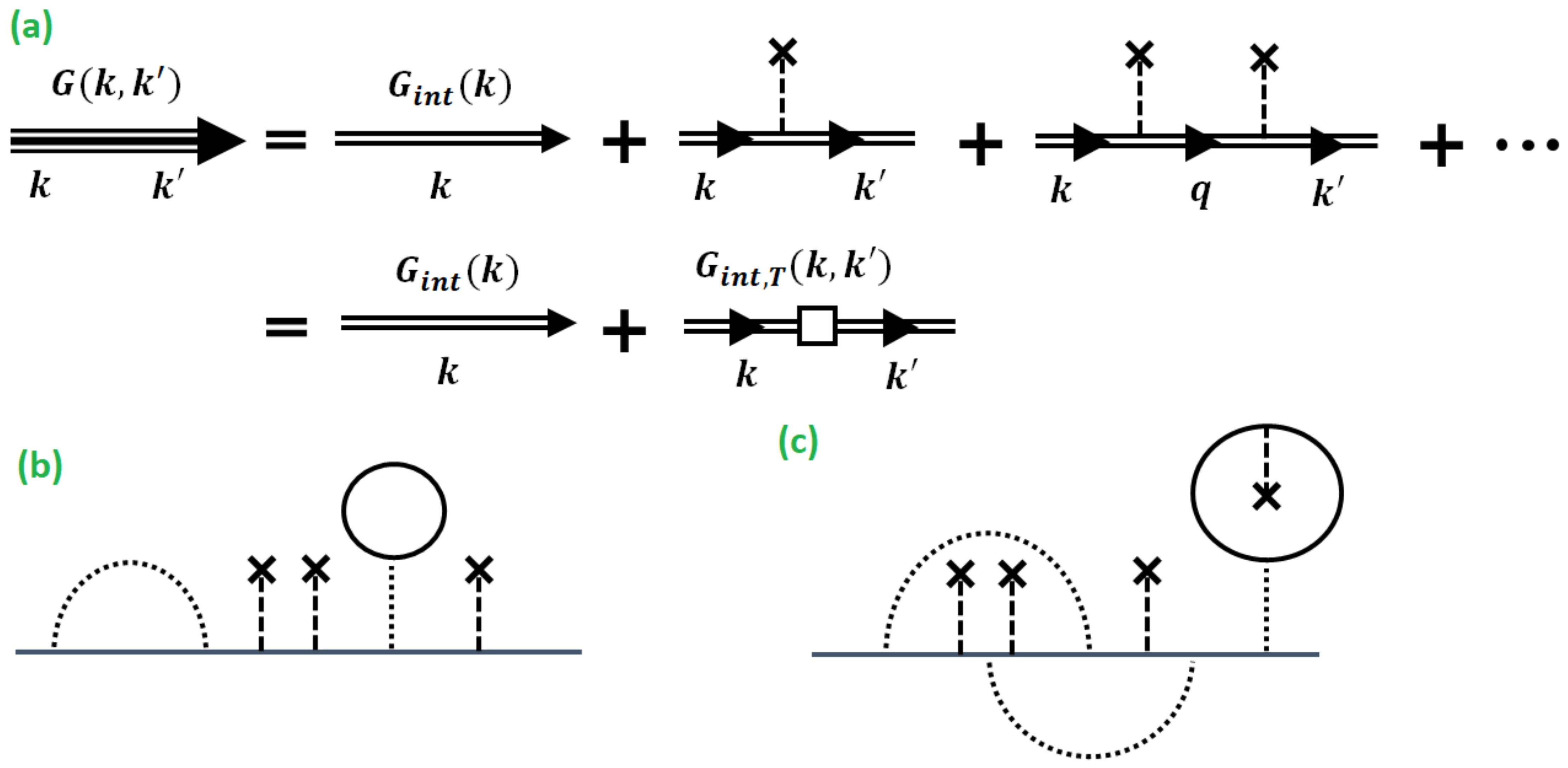}
\caption{ (a) Graphical presentation of the full impurity Green's function solved by incorporating the homogeneous interacting Green's function into the $T$-matrix formalism. (b) Examples of the diagrams that are (b) included and (c) not included in our approach. } 
\label{fig:Gint_imp_Feynmann_diagrams}
\end{center}
\end{figure}

As mentioned above, the presence of edge states is achieved through $T$-matrix formalism by the presence of an impurity that simulates an edge. In this work we assume a sharp $\delta$-function to model the edge (impurity) located at $r=0$  
\begin{eqnarray}
U(r)=U_{0}\delta(r)\;,
\label{Ur_delta_fn}
\end{eqnarray}
for a 1D TI defined in the $r>0$ semi-infinite space. For a general potential scattering, with strength in momentum space denoted by $U_{kk'}$, the full Green's function is no longer translational invariant, and hence in momentum space it reads
\begin{eqnarray}
G(k,k',\tau)=\left(
\begin{array}{ll}
G_{AA}(k,k',\tau) & G_{AB}(k,k',\tau) \\
G_{BA}(k,k',\tau) & G_{BB}(k,k',\tau)
\end{array}
\right)\;,
\label{full_GF_kkp}
\end{eqnarray}
with the matrix elements defined by 
\begin{eqnarray}
G_{IJ}(k,k',\tau)=-\langle T_{\tau}c_{Ik}(\tau)c_{Jk'}^{\dag}(0)\rangle\;.
\end{eqnarray}
In the presence of interactions, the impurity (or "edge") Green's function, denoted by $G$, can be obtained by incorporating the $G_{int}$ in Eq.~(\ref{2D_G_dprime}) into the $T$-matrix approximation\cite{Balatsky06}
\begin{eqnarray}
G(k,k')&=&G_{int}(k)+G_{int}(k)U_{kk'}G_{int}(k')+...
\nonumber \\
&=&G_{int}(k)+G_{int}(k)T_{kk'}G_{int}(k')\;,
\nonumber \\
&=&G_{int}(k)+G_{int-T}(k,k')\;,
\label{impurity_GF}
\end{eqnarray}
where $G_{int}(k)$ and $G_{int-T}(k,k')$ denote, respectively, the homogeneous (no impurity) and inhomogeneous (with impurity) parts. The $T$-matrix satisfies 
\begin{eqnarray}
T_{kk'}=U_{kk'}+\sum_{k''}U_{kk''}G_{int}(k'')T_{k''k'}\;,
\end{eqnarray}
as shown in Fig.~\ref{fig:Gint_imp_Feynmann_diagrams}.
For the $\delta$-function potential we used in Eq.~(\ref{Ur_delta_fn}) to simulate the edge, one has $U_{kk'}=U_{0}$, and hence
\begin{eqnarray}
T=T(i\omega_{n})=U_{0}\left[I-U_{0}\sum_{k}G_{int}(k,i\omega_{n})\right]^{-1}\;.
\end{eqnarray}
We see that the $T$-matrix is only a function of the (Matsubara) frequency. Particularly in the hard edge limit $U_{0}\rightarrow\infty$, the $T$-matrix becomes 
\begin{eqnarray}
\lim_{U_{0}\rightarrow\infty}T(i\omega_{n})=-\left[\sum_{k}G_{int}(k,i\omega_{n})\right]^{-1}\;,
\end{eqnarray} 
which takes a rather simple form. The $T$-matrix with real frequencies, which is required to calculate the full retarded Green's function mentioned below, is evaluated in the same manner. This method takes into account the essential diagrams that captures the edge state, i.e., the diagrams that produce the impurity bound state. However, it also does not take into account a variety of impurity-interaction interference diagrams as shown in Fig.~\ref{fig:Gint_imp_Feynmann_diagrams} (c), which should be addressed elsewhere.

After solving the full retarded Green's function, equivalent to replacing $i\omega_{n}\rightarrow\omega+i\eta$ everywhere in the formalism, we now have all the ingredients to obtain the LDOS. We first Fourier transform the momentum space full Green's function into real space full Green's function 
\begin{eqnarray}
G^{ret}_{IJ}(r,r',\omega)=\int\frac{dk}{2\pi}\int\frac{dk'}{2\pi}e^{i(kr-k'r')}G^{ret}_{IJ}(k,k',\omega)\;.
\nonumber \\
\label{Grrp_Fourier}
\end{eqnarray}
The real space LDOS
\begin{eqnarray}
\rho(r,\omega)&=&-{\rm Im}\left\{{\rm Tr}\left[G^{ret}(r,r,\omega)\right]\right\}/\pi
\nonumber \\
&=&-{\rm Im}\left\{G^{ret}_{AA}(r,r,\omega)+G^{ret}_{BB}(r,r,\omega)\right\}/\pi\;.
\label{LDOS}
\end{eqnarray}
comprises the sum of imaginary parts of the retarded Greens function of the $A$ and $B$ sublattices. The spectral sum rule of this $T$-matrix-Dyson's equation formalism is satisfied in the following manner. If we separate the homogeneous part and the impurity part of the Green's function as in Eq.~(\ref{impurity_GF}), and denote their corresponding LDOS as 
\begin{eqnarray}
&&\rho_{int}(r,\omega)=-{\rm Im}\left\{{\rm Tr}\left[G_{int}^{ret}(r,r,\omega)\right]\right\}/\pi\;,
\nonumber \\
&&\rho_{int-T}(r,\omega)=-{\rm Im}\left\{{\rm Tr}\left[G_{int-T}^{ret}(r,r,\omega)\right]\right\}/\pi\;,
\nonumber \\
&&\rho(r,\omega)=\rho_{int}(r,\omega)+\rho_{int-T}(r,\omega)\;,
\end{eqnarray}
then 
\begin{eqnarray}
&&\int_{-\infty}^{\infty} d\omega\, \rho(r,\omega)=\int_{-\infty}^{\infty} d\omega\, \rho_{int}(r,\omega)=2\;,
\nonumber \\
&&\int_{-\infty}^{\infty} d\omega\, \rho_{int-T}(r,\omega)=0\;.
\end{eqnarray}
That is, the particle number is conserved due to the homogeneous part (the integration gives $2$ because we count both the conduction band and the valance band). The impurity-scattering part $\rho_{int-T}(r,\omega)$ is what gives the edge state (see following sections), but it only modifies the profile of the total LDOS $\rho(r,\omega)$ and does not change total number of particles, as expected since $\left[H,H_{e-e}\right]=0$.


To accurately extract the decay length of the edge state, we use the following method without the tedious fit of the decaying LDOS in real space. Near the critical point $M\rightarrow M_{c}$, the impurity Green's function in momentum space $G^{ret}(k,k',\omega)$ is even in momentum space and well fits a Ornstein-Zernike form near the high symmetry points $\left\{k_{0},k'_{0}\right\}=\left\{\pi,\pi\right\}$, i.e.,
\begin{eqnarray}
G^{ret}(k_{0}+\delta k,k'_{0}+\delta k',\omega)\approx\frac{G^{ret}(k_{0},k'_{0},\omega)}{(1+\xi^{2}\delta k^{2})(1+\xi^{2}\delta k'^{2})}.\;\;\;\;\;
\label{Gkkp_Ornstein_Zernike}
\end{eqnarray} 
Consequently, after the Fourier transform in Eq.~(\ref{Grrp_Fourier}), the LDOS decays with a decay length $\xi$. Thus we only need to extract $\xi$ from the Ornstein-Zernike fit in Eq.~(\ref{Gkkp_Ornstein_Zernike}), which is a much simpler task than explicitly performing the Fourier transform in Eq.~(\ref{Grrp_Fourier}) and then fit $\xi$ in real space.

\section{Su-Schrieffer-Heeger model with interactions}

\subsection{Topological invariant and edge states of interacting SSH model}

\begin{figure}[ht]
\begin{center}
\includegraphics[clip=true,width=0.99\columnwidth]{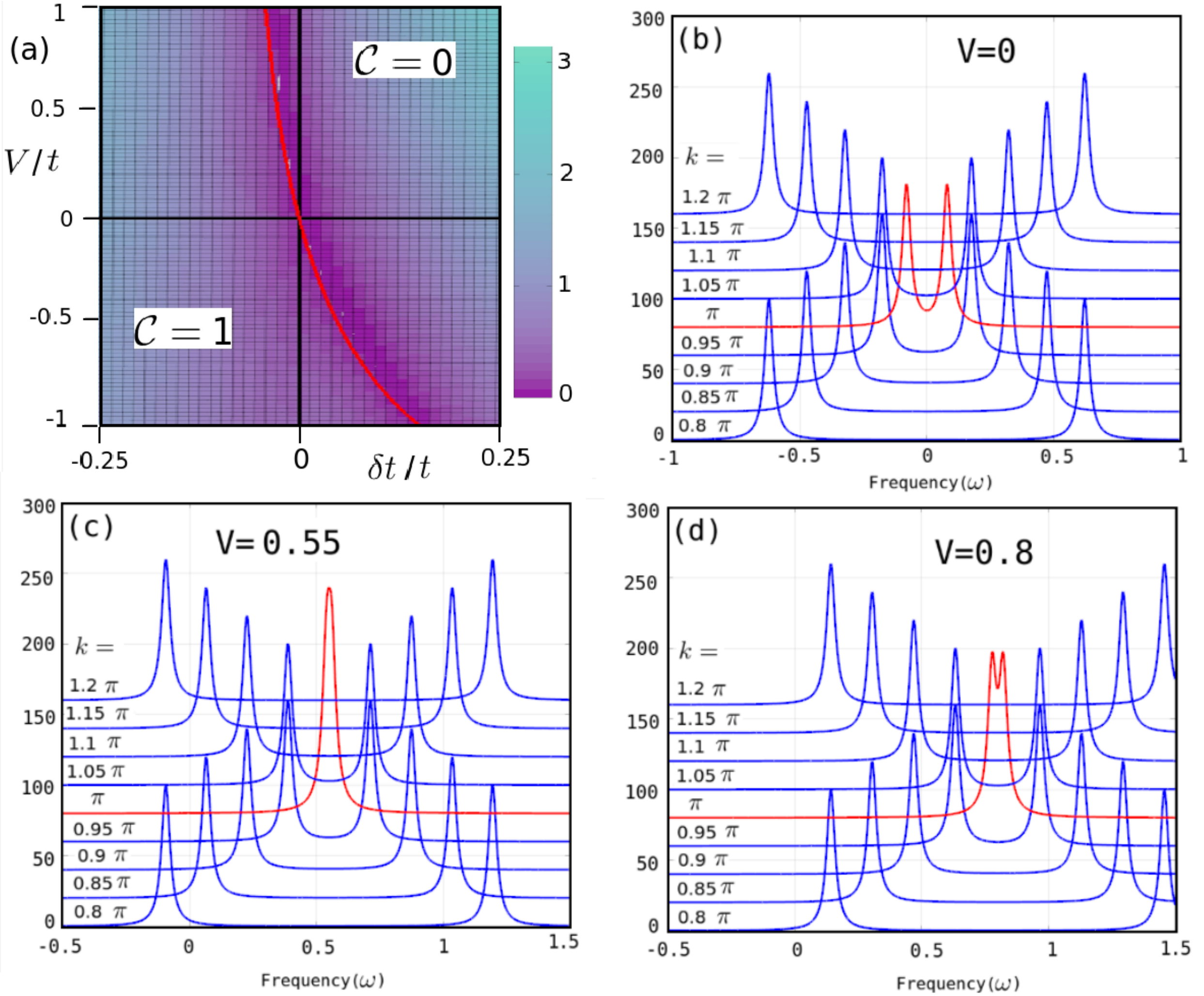}
\caption{ (a) Phase diagram of the SSH model with nearest-neighbor interaction. The color scale indicates the size of the gap identified from the spectral function $A(k,\omega)$. (b) to (d) show $A(k,\omega)$ as a function of $\omega$ from $k=0.8\pi$ to $1.2\pi$ at $\delta t=-0.04$ at different values of $V$. The gap-closing at the critical point $V_{c}=0.55$ is evident. The entire spectrum shifts with $V$ because the interaction in Eq.~(\ref{SSH_ee_interaction}) breaks particle-hole symmetry. } 
\label{fig:SSH_Akw_gap_closing}
\end{center}
\end{figure}

As a concrete example of the approach described in the previous sections, here we consider the spinless Su-Schrieffer-Heeger (SSH) model in the presence of electron-electron interaction. Various versions of interacting SSH model has been studied previously and investigated by different methods\cite{Marques17,Marques18,Yahyavi18,Sbierski18,Kuno19}. The topological properties in the noninteracting limit, on the other hand, has been studied experimentally by means of optical lattices\cite{Atala13,Meier16}. Here we focus on the quantum criticality investigated by means of the Green's function formalism in Secs.~\ref{sec:Green_fn_Dirac_interaction} and \ref{sec:edge_state_Dirac}. The noninteracting part of the Hamiltonian is 
\begin{eqnarray}
{\cal H}_{0}&=&\sum_{i}(t+\delta t)c_{Ai}^{\dag}c_{Bi}+(t-\delta t)c_{Ai+1}^{\dag}c_{Bi}+h.c.
\nonumber \\
&=&\sum_{k}Q_{k}c_{Ak}^{\dag}c_{Bk}+Q_{k}^{\ast}c_{Bk}^{\dag}c_{Ak}\;,
\label{SSH_unperturbed}
\end{eqnarray}
where $t+\delta t$ and $t-\delta t$ are the hopping amplitudes on the even and the odd bonds, respectively, and $Q_{k}=(t+\delta t)+(t-\delta t)e^{-ik}$. Here we consider the nearest-neighbor interaction as in Ref.~\onlinecite{Chen18}
\begin{eqnarray}
{\cal H}_{e-e}=V\sum_{i}\left(n_{Ai}n_{Bi}+n_{Bi}n_{Ai+1}\right)\;,
\label{SSH_ee_interaction}
\end{eqnarray}
where $n_{Ii}\equiv c_{Ii}^{\dag}c_{Ii}$. The Fourier transform of the above equation yields Eq.~(\ref{general_density_density_interaction}) with $V_{q}=V(1+\cos q)$~\cite{Chen18}. Notice that this form of interaction breaks particle-hole and chiral symmetries with respect to zero frequency (the ${\mathcal C}$ and ${\mathcal S}$ symmetric version is given by replacing $n_{Ii}\rightarrow(n_{Ii}-1/2)$ everywhere in Eq.~(\ref{SSH_ee_interaction})). Up to one-loop approximation, the self-energies are given by~\cite{Chen18}
\begin{eqnarray}
&&\Sigma_{AA}(k)=\Sigma_{BB}(k)=V\;
\label{SSH-off-self-energy},\\ 
&&\Sigma_{AB}(k)=\frac{1}{2}\sum_{q}V_{q}e^{-i\alpha_{k+q}}
=\left[\Sigma_{BA}(k)\right]^{\ast}\;,
\label{SSHnn_self-energy}
\end{eqnarray}
where the phase $\alpha_k$ is defined from $Q_{k}\equiv|Q_{k}|e^{-i\alpha_{k}}$. The $\Sigma_{AA}$ and $\Sigma_{BB}$ are the Hartree terms that introduce a finite chemical potential $d_{0}'=-V$ that shifts the entire spectrum by $V$. Using Eq.~(\ref{newC}), it becomes clear that 
\begin{eqnarray}
{\mathcal S}\left[H\left(k\right)+V\right]{\mathcal S}^{-1} & =-\left[H\left(k\right)+V\right],\label{newS2}\\
{\mathcal C}\left[H\left(-k\right)+V\right]{\mathcal C}^{-1} & =-\left[H\left(k\right)+V\right],\label{newC2}
\end{eqnarray}
which generalizes the symmetries with respect to the finite chemical potential, thus yielding a spectrum that is symmetric with respect to $V$. Finally, using Eqs.~(\ref{SSH-off-self-energy}) and (\ref{SSHnn_self-energy}), the Green's function Eq.~(\ref{2D_G_dprime}) becomes 
\begin{eqnarray}
&&G(k,i\omega)=\frac{1}{(i\omega-V)^{2}-|Q_{k}+\Sigma_{AB}(k)|^{2}}
\nonumber \\
&&\times\left(
\begin{array}{cc}
i\omega-V & Q_{k}+\Sigma_{AB}(k) \\
\left[Q_{k}+\Sigma_{AB}(k)\right]^{\ast} & i\omega-V
\end{array}
\right)\;.
\end{eqnarray}
As discussed in Sec.~\ref{sec:Green_fn_Dirac_interaction}, at $i\omega=V$ the full Green's function becomes off-diagonal. The topological invariant corresponds to the winding number of the phase\cite{Essin11,Chen18,Sbierski18} 
\begin{eqnarray}
&&\varphi_{k}=-\arg\left(Q_{k}+\Sigma_{AB}\right),
\nonumber \\
&&=-\arg\left(Q_{k}+\frac{1}{2}\sum_{q}V_{q}e^{-i\alpha_{k+q}}\right)\;.
\label{SSHnn_phik_analytical_result}
\end{eqnarray}
The topological phase transition in the $\delta t-V$ parameter space, identified from the gap-closing in the spectral function (see discussion below), coincides with that solved by the CRG approach in Ref.~\cite{Chen18}, as shown in Fig.~\ref{fig:SSH_Akw_gap_closing} (a). The phase diagram shows both topologically trivial ${\cal C}=0$ and nontrivial ${\cal C}=1$ phases in  either $V>0$ or $V<0$ regimes. We note, however, that for $V<0$, the topological phase transition has moved from $\delta t=0$ to a negative $\delta t$, while for $V>0$ it moved to a positive $\delta t$. Therefore, the presence of the interaction does change the critical point in the parameter space.


\begin{figure}[ht]
\begin{center}
\includegraphics[clip=true,width=0.99\columnwidth]{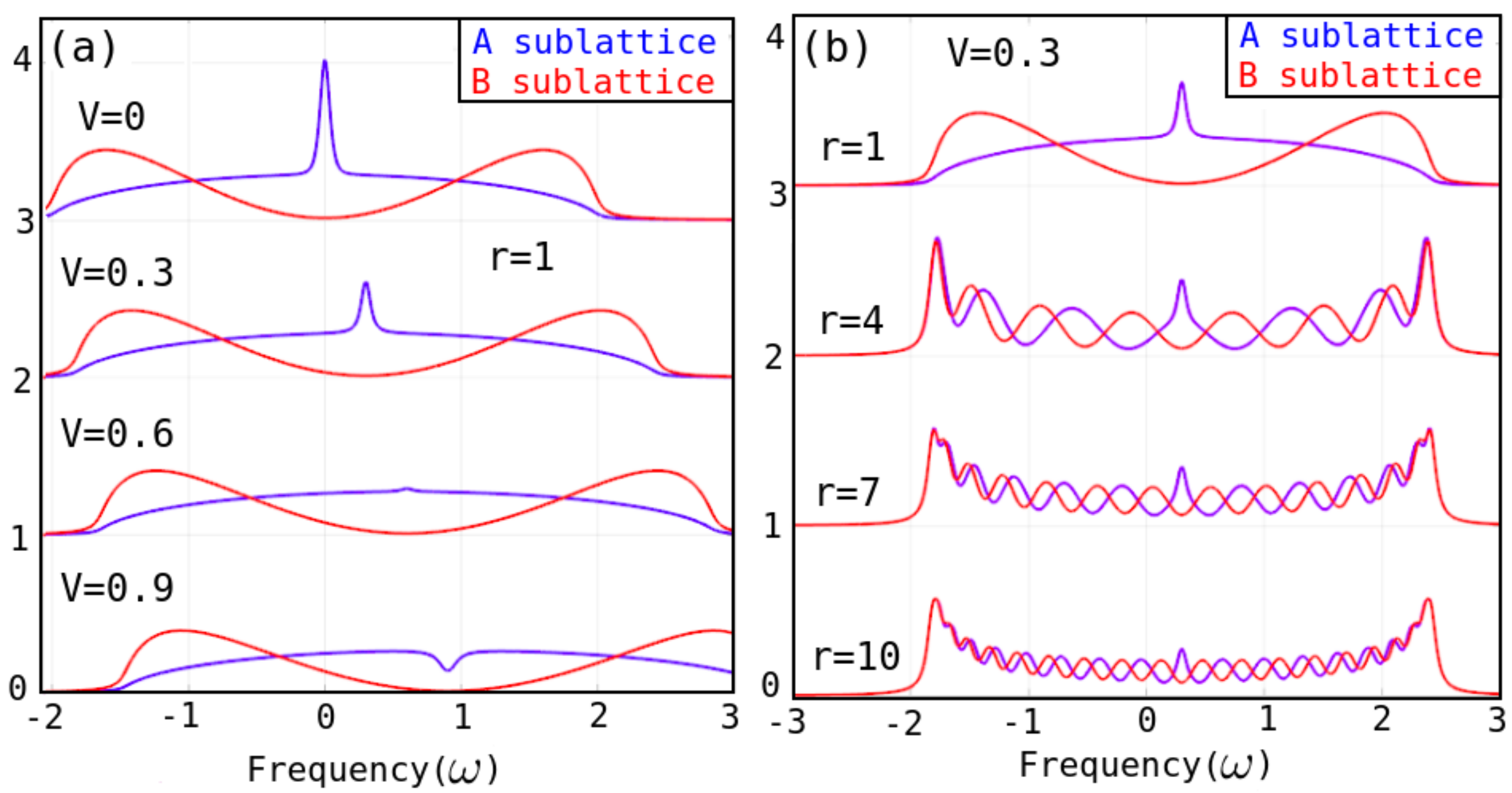}
\caption{ (a) LDOS on the two sublattices of SSH model with $\delta t=-0.04$ at the first site $r=1$ away from the edge, and for different values of interaction $V$. One sees that the edge state in the middle of the spectrum only exists in the topologically nontrivial phase $V\apprle 0.6$, and is only localized in the $A$ sublattice. (b) LDOS for $V=0.3$ at different sites $r$ away from the edge, which manifests an edge state that decays with $r$. The wavy features in the spectrum are due to finite size effects.  } 
\label{fig:SSH_LDOS_vsV_vsr}
\end{center}
\end{figure}

\begin{figure}[ht]
\begin{center}
\includegraphics[clip=true,width=0.99\columnwidth]{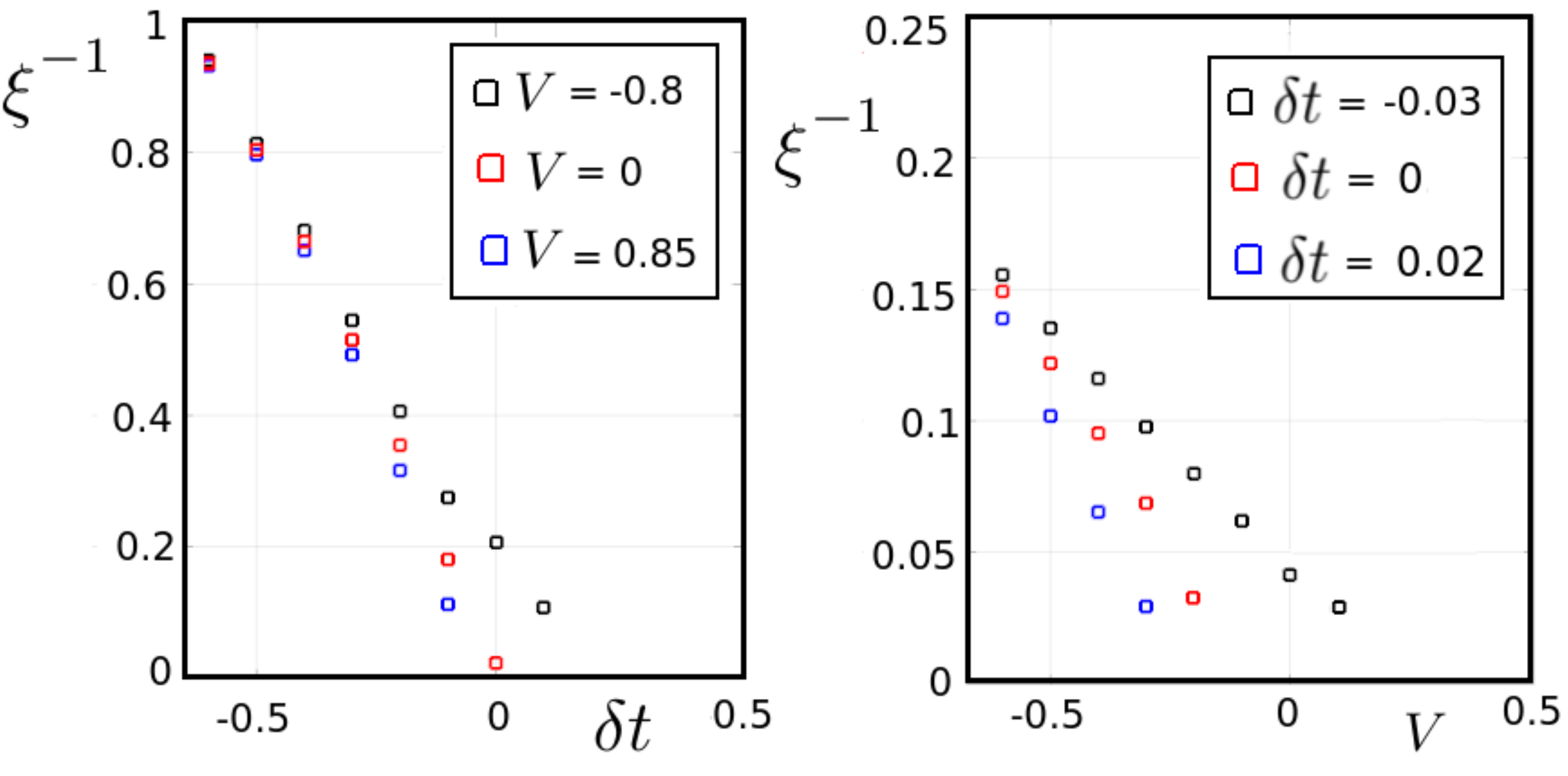}
\caption{ Fitting the critical exponent of the decay length of the edge state $\xi\sim|M|^{-\nu}$ for (a) $M=\delta t$ at several values of $V$, and (b) $M=V$ at several values of $\delta t$. } 
\label{fig:SSH_xi_critical_exponent}
\end{center}
\end{figure}


The spectal function $A(k,\omega)$ calculated from Eq.~(\ref{spectral_fn_Gint}) is shifted in frequency by the interaction $V$ because of the particle-hole-breaking nature of the interaction, as shown in Fig.~\ref{fig:SSH_Akw_gap_closing} (b)$\sim$(d). To be more specific, the shift comes from the Hartree term $\Sigma_{AA}(k)=\Sigma_{BB}(k)=V$ in Eq.~(\ref{SSHnn_self-energy}). In addition, because the Hartree term is entirely real, it does not influence the quasiparticle lifetime, i.e., the spectral function is not broadened by the electron-electron interaction we considered. Despite the shift in frequency at finite $V$, the band gap, identified from the distance between the two peaks $A(\pi,\omega)$ at the high symmetry point $k=\pi$, is clearly visible in Fig.~\ref{fig:SSH_Akw_gap_closing} (b)$\sim$(d). As $V$ is tuned across t critical point $V_{c}$, the spectral function clearly displays a gap-closing at the high symmetry point. For the topological phase transitions driven by the kinetic parameter $\delta t$, the same thing happens. Thus the spectral function unambiguously demonstrates that gap-closing in a defining feature for topological phase transitions even in interacting systems.

The LDOS in this interacting SSH model is shown in Fig.~\ref{fig:SSH_LDOS_vsV_vsr}. The edge state is identified from the peak in the middle of the spectrum, which only appears in the topologically nontrivial phase, indicating that the bulk-edge correspondence is well satisfied in this model. Because the entire spectrum is shifted by the interaction $V$ and the edge state remains in the middle of the gap, the edge state is shifted away from absolute zero frequency by the interaction. As a function of the distance away from the edge at $r=0$, we uncover that the edge state remains localized only in one sublattice, a feature inherited from the noninteracting limit which can be obtained by analytically solving the Dirac equation projected into real space (see, for instance, Ref.~\onlinecite{Chen17}). The decay of the edge state with $r$ is clearly visible, with a decay length $\xi\sim $M$^{-1}$ that exhibits critical exponent $\nu=1$ regardless whether the phase transition is driven by the kinetic parameter $M=\delta t$ or the interaction $M=V$, as shown in Fig.~\ref{fig:SSH_xi_critical_exponent}. This indicates that the SSH model, despite in the presence of the nearest-neighbor interaction in Eq.~(\ref{SSH_ee_interaction}), remains in the same universality class.

\section{Conclusions}

In summary, we present a Green's function formalism to investigate the bulk-edge correspondence and the quantum criticality in weakly interacting 1D TIs. Using SSH model with nearest-neighbor interaction as an example, we reveal a spectral function that indicates gap-closing is still a defining feature even for interaction-driven topological phase transitions. By employing the $T$-matrix formalism, the edge state can be identified from the LDOS near the impurity (edge) site, and it is found to locate in the same sublattice as the noninteracting edge state. The entire energy spectrum is shifted by the Hartree term of the self-energy if the interaction breaks particle-hole symmetry, and because the edge state remains in the middle of the gap, it can be shifted away from zero energy by the interaction. Extracting the critical exponent from the decay length of the edge state reveals that the system remains in the same universality class even in the presence of interactions, at least within our approach that is limited within one-loop calculation and the $T$-matrix approximation that neglects the interference diagrams.

We anticipate that our formalism can be generalized to address similar issues for TIs and TSCs in different dimensions and symmetry classes, as well as various different kinds of interactions. The edge state in higher dimensions may be simulated by a line or a surface of impurities in the $T$-matrix formalism. As the self-energy caused by electron-electron interaction investigated in this article remains frequency-independent and hence does not influence the quasiparticle lifetime, it will be enlightening to investigate other types of self-energies that are frequency-dependent, for instance those arising from electron-phonon interactions. In such cases, how the broadened spectral functions influence the gap-closing could be an interesting subject. Moreover, systems belonging to different symmetry classes may be properly simulated by choosing the appropriate minimal Dirac models. All these perspectives await further investigations. An interesting direction would be to investigate the robustness of the topological and non-topological edges states in quantum dots of Ref.~\cite{candido2018prl} against the electron-electron scattering studied here. Clearly additional work is needed to elucidate the whether our results reported here would hold in confined geometries such as quantum dots.

\section{Acknowledgement}

WC acknowledges the financial support from the incentives to research productivity fellowship from PUC-Rio, and the productivity in research fellowship from CNPq. JCE and DRC acknowledge support from the S\~{a}o Paulo Research Foundation (FAPESP grant No. 2016/08468-0), CNPq, and PRP/USP (QNANO).

\bibliography{Literatur}

\end{document}